\documentclass[twocolumn,amsmath,amssymb,pra,aps]{revtex4-2}

\usepackage{graphicx}
\usepackage{physics}
\usepackage{dcolumn}% Align table columns on decimal point
\usepackage{bm}% bold math
\usepackage{ragged2e}
\usepackage{float}

\newcommand{\vect}[1]{\boldsymbol{\mathbf{#1}}}

\usepackage{hyperref}% add hypertext capabilities
\hypersetup{colorlinks=true,citecolor=blue,linkcolor=blue,urlcolor=blue}
\newcommand{\rmsss}[1]{\mathrm{\scriptscriptstyle{#1}}}

\newcommand{\kN}{k_{\rmsss{N}}}
\newcommand{\kNz}{k_{\rmsss{N}_z}}

\newcommand{\vN}{v_{\rmsss{N}}}
\newcommand{\phiN}{\varphi_{\rmsss{N}}}

\newcommand{\kS}{k_{\rmsss{S}}}
\newcommand{\kSz}{k_{\rmsss{S}_z}}
\newcommand{\vSz}{v_{\rmsss{S}_z}}
\newcommand{\vS}{v_{\rmsss{S}}}
\newcommand{\phiS}{\varphi_{\rmsss{S}}}

\newcommand{\kF}{\kappa_\rmsss{F}}

\newcommand{\mN}{m_{\rmsss{N}}}

\newcommand{\mS}{m_{\rmsss{S}}}

\newcommand{\dN}{d_{\rmsss{N}}}
\newcommand{\dS}{d_{\rmsss{S}}}
\newcommand{\dF}{d_{\rmsss{F}}}

\newcommand{\tN}{\tau_{\rmsss{N}}}
\newcommand{\tS}{\tau_{\rmsss{S}}}

\begin{document}

\title{Topological superconductivity in tripartite superconductor-ferromagnet-semiconductor nanowires}

\author{Josias Langbehn}
\author{Sergio Acero Gonz\' alez}
\author{Piet W. Brouwer}
\author{Felix von Oppen}

\affiliation{Dahlem Center for Complex Quantum Systems and Fachbereich Physik, Freie Universit\"at Berlin, 14195 Berlin, Germany }

\date{\today}

\begin{abstract}
Motivated by recent experiments searching for Majorana zero modes in tripartite semiconductor nanowires with epitaxial superconductor and ferromagnetic-insulator layers, we explore the emergence of topological superconductivity in such devices for paradigmatic arrangements of the three constituents. Accounting for the competition between magnetism and superconductivity, we treat superconductivity self consistently and describe the electronic properties, including the superconducting and ferromagnetic proximity effects, within a direct wave-function approach. We conclude that the most viable mechanism for topological superconductivity relies on a superconductor-semiconductor-ferromagnet arrangement of the constituents, in which spin splitting and superconductivity are independently induced in the semiconductor by proximity and superconductivity is only weakly affected by the ferromagnetic insulator.  
\end{abstract}

\maketitle

\section{Introduction}

Topological superconductors can be engineered through a combination of spin-orbit coupling, conventional superconductivity, and Zeeman splitting \cite{AliceaReview,BeenakkerReview,LutchynReview}. A candidate platform are hybrid semiconductor-superconductor nanowires pierced by an external magnetic field \cite{Lutchyn2010,Oreg2010}. The semiconductor, typically InAs or InSb, provides the spin-orbit coupling, while the superconductor and the magnetic field contribute the conventional superconductivity and the Zeeman splitting, respectively. Experiments using this scheme have reported the observation of zero-bias peaks, consistent with the presence of Majorana zero modes \cite{Mourik2012,Rokhinson2012,Deng2012,Das2012,Deng2016,Suominen2017, Chen2017,Zhang2017,Gul2018,Lutchyn2018,Deng2018,Manna2020,Liu2019, Vaitieknas2020,LutchynReview}. 
Even if this blueprint proved consistently successful for the engineering of topological superconductivity, the use of an external magnetic field might be inconvenient for engineering more involved devices underlying a Majorana-based quantum computer \cite{OregOppenAnnualReview}. In particular, the magnetic field should ideally be applied parallel to the nanowire, requiring all nanowires to be aligned. 

In an effort to alleviate this constraint, recent experiments \cite{Liu2019, Vaitieknas2020,Manna2020} have explored the possibility of replacing the external magnetic field by a proximity-induced exchange field exerted by an epitaxial ferromagnetic insulator grown directly on the nanowire. One set of experiments \cite{Liu2019,Vaitieknas2020} uses semiconductor nanowires (InAs) with epitaxial superconducting (Al) and ferromagnetic (EuS) layers. Another experiment \cite{Manna2020} grows Au wires on top of a superconducting substrate (V) and covers them by a EuS layer. Motivated by these experiments, we study the emergence of topological superconductivity in such tripartite nanowires, which combine a semiconducting or metallic core (N) with epitaxial superconducting (SC) and ferromagnetic (F) layers, from a theoretical perspective, complementing a series of concurrent studies \cite{Woods2020a,Escribano2020,Maiani2020,Liu2020}.

\begin{figure}[t]
\centering
\includegraphics[width=0.9\columnwidth]{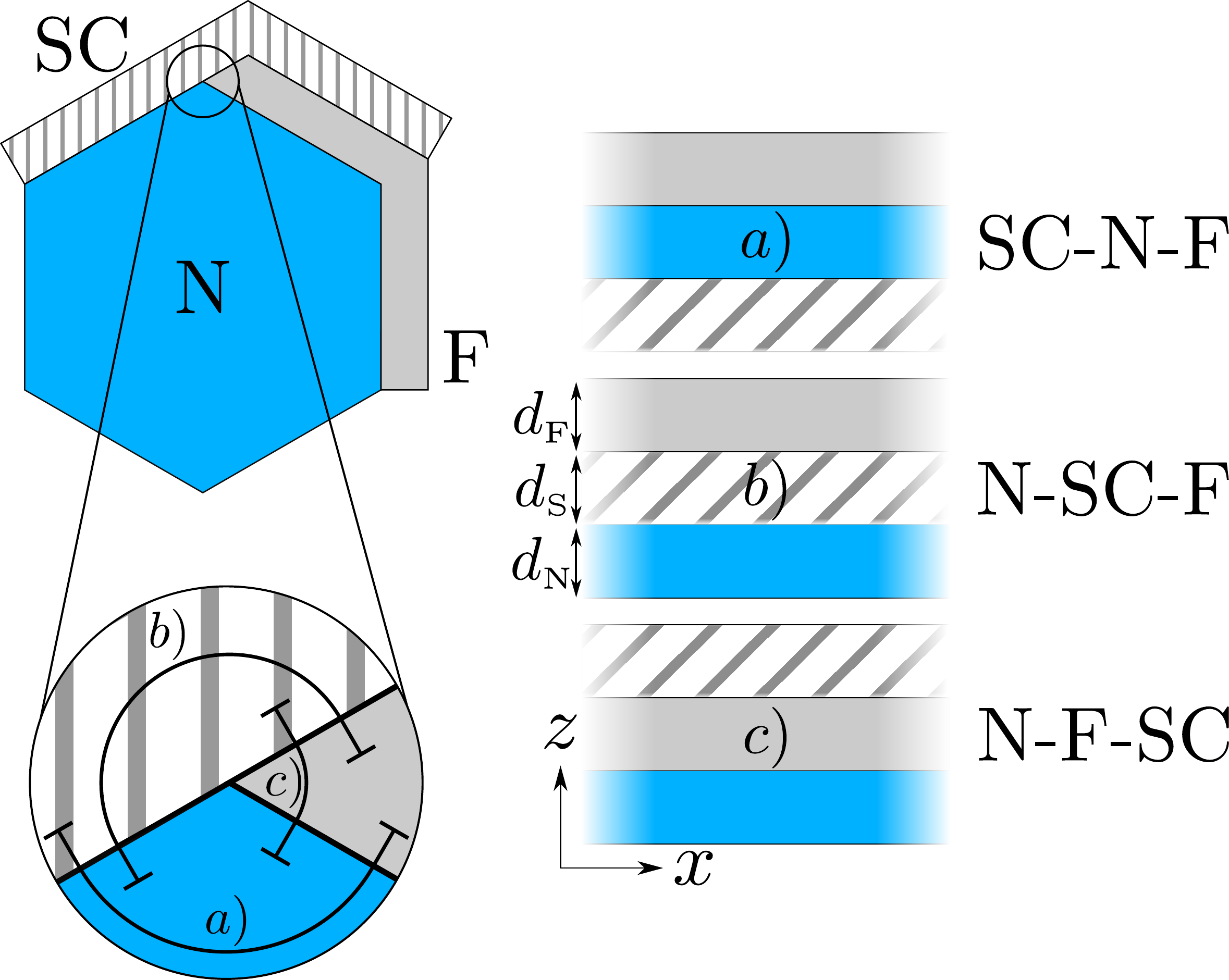}
\caption{Top left: Schematic representation of the nanowire geometry (cross section) employed in the experiments in Refs.\ \cite{Liu2019,Vaitieknas2020}. Bottom left: Enlarged view of the region, presumably most important for the emergence of topological supercondutivity, where the semiconductor (N), the superconductor (SC), and the ferromagnetic insulator (F) meet. Right: Three paradigmatic stackings of N, SC, and F, which we investigate to explore the emergence of topological superconductivity: (a) SC-N-F, (b) N-SC-F, and (c) N-F-SC. In experiment, the diameter of the nanowire is of the order of $100$ nm, with epitaxial SC and F layers of thickness $\sim 5$nm. \label{fig:hexagonal_wire}}
\end{figure}

A schematic section through the experimental nanowires in Refs.\ \cite{Liu2019,Vaitieknas2020} is shown in Fig.\ \ref{fig:hexagonal_wire}. A semiconducting nanowire with hexagonal cross section is covered by a ferromagnetic insulator on one facet. The superconducting layer covers both a neighboring facet as well as the ferromagnetic layer. Band bending at the normal-superconductor interface is expected to lead to electron accumulation near that interface, presumably making the region where all three layers meet particularly pertinent for the potential emergence of topological superconductivity. As seen from the enlarged rendering in Fig.\ \ref{fig:hexagonal_wire}, this region includes interfaces between all three layers. Exemplifying the experimental geometries by stacks of three layers, we thus study the emergence of topological superconductivity for the three possible stackings as shown in Fig.\ \ref{fig:hexagonal_wire}. Such a stacked structure also closely resembles the experimental setup in Ref.\ \cite{Manna2020}. 

We describe both the ferromagnetic and the superconducting proximity effects underlying the emergence of topological superconductivity in these structures within a direct wave-function approach. For stackings involving an interface between the superconductor and the ferromagnetic insulator, the superconducting pairing will be substantially suppressed. We account for this competition by determining the superconducting pairing self-consistently. The ferromagnetic proximity effect on a thin superconducting layer resembles, but is not identical to the effect of an external Zeeman field \cite{Meservey1970,Tedrow1986,Hao1991,Wolf2014,Strambini2017}, and is uniform across the entire superconducting layer as long as its thickness is small compared to the superconducting  coherence length \cite{Tokuyasu1988,Bergeret2004}. We find that all three possible layer arrangements can support topological superconductivity. However, the effects of the ferromagnet on the superconductor greatly limit the extent of the topological phase in parameter space, when a direct SC-F interface is present. Moreover, a ferromagnetic insulator sandwiched between superconductor and semiconductor will tend to decouple the semiconductor from the superconductor, so that a possible topological superconducting phase occurs only for very thin F layers.  We thus find that the topological superconducting phase has the largest extent in parameter space for the SC-N-F arrangement, where the emergence of the topological phase closely parallels the familiar blueprint \cite{Lutchyn2010,Oreg2010}. 

We begin with a physical discussion in Sec.\ \ref{sec:phys_disc}, where we provide semiclassical estimates and present the main results of our work. In Sec.\ \ref{sec:model}, we detail our model and the numerical calculations, including the self-consistent treatment of superconductivity. In Sec.\ \ref{sec:Phase_Diags}, we elaborate on the phase diagrams, which we obtain numerically. Finally, we conclude in Sec.\ \ref{sec:conclusion}.\par

\section{Physical picture} \label{sec:phys_disc}

This section provides a summary of our principal results on the basis of physical arguments. We begin with a brief discussion of the proximity effect induced by a ferromagnetic insulator. When a ferromagnetic insulator is brought into contact with a normal metal or a superconductor, it induces a spin polarization of the carriers. Carriers impinging on the interface with the ferromagnet are reflected, with the penetration depth into the ferromagnet depending on their spin state. This spin-dependent penetration reflects the different band gaps for the two spin projections and is reflected in spin-dependent scattering phases. In a semiclassical picture (Bohr-Sommerfeld quantization), it is evident that this makes the subband energies spin dependent, effectively inducing a spin splitting analogous to a Zeeman field.  

\begin{figure*}[t]
\centering
\includegraphics[width=\textwidth]{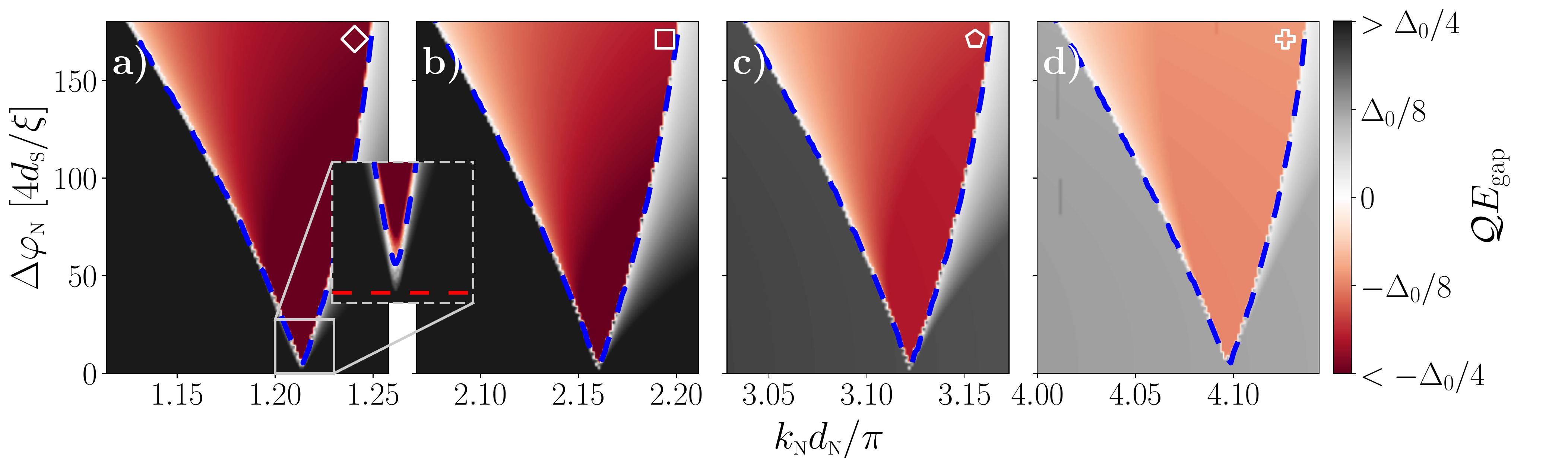}
\caption{Phase diagram for the {SC-N-F} arrangement. The magnitude of the overall gap $E_\mathrm{gap}$, multiplied by the topological invariant $\mathcal{Q}= \pm 1$, is color coded as a function of the phase difference $\Delta \phiN$ and the number $\kN\dN/\pi$ of occupied transverse modes in the normal layer. The four panels (a)-(d) focus on those regions, where the first four modes of the N layer begin to be populated. Phase-transition lines are indicated by blue dashed lines. The inset in panel (a) enlarges the apex of the topological region, and includes the semiclassical estimate Eq.\ (\ref{eq:phasediff_estimate}) for the minimal spin-dependent phase difference required  to enter the topological superconducting phase (red dashed line). The symbol in the top right corner of the panels indicates the parameter choice as shown in Fig.\ \ref{fig:resonance_all}. 
The number of transverse modes in S was fixed at $\kS\dS/\pi = 27.52$ for all panels. For other parameters, see Table \ref{tab:real_parameters}.
\label{fig:SNF_phase_diagram}}
\end{figure*}

While this proximity-induced spin splitting is closely analogous to the effects of a Zeeman field, there are also characteristic differences. To appreciate these differences, compare the effects of a Zeeman field and a proximitizing ferromagnetic insulator on a thin-film superconductor. With increasing Zeeman field, the normal state becomes magnetized and energetically more favorable. Beyond the Clogston-Chandrasekhar limit \cite{Clogston1962,Chandrasekhar1962}, the energy gain due to the magnetization is larger than the superconducting condensation energy, resulting in a first-order phase transition between the superconducting and normal states. In contrast, when proximity coupling the thin-film superconductor to a ferromagnetic insulator, one expects a second-order phase transition when increasing the exchange field exerted by the ferromagnet \cite{Tokuyasu1988}. The underlying reason is that the effect of the ferromagnet depends on the transverse mode in the superconductor, even when the thickness of the thin film is small compared to the superconducting coherence length and the superconductor becomes uniformly magnetized \cite{Tokuyasu1988,Bergeret2004,Strambini2017}. Semiclassically, the transverse modes can be thought of as electron trajectories impinging on the interface with the ferromagnetic insulator at mode-specific angles. Due to Andreev reflection, the overall length of the trajectory in the superconductor is limited by the superconducting coherence length $\xi = \hbar\vS/\Delta_0$. (Here, $\vS$ is the Fermi velocity in the superconductor and $\Delta_0$ the unperturbed superconducting pairing). Thus, for a superconductor of thickness $\dS$, the trajectories reflect from the ferromagnet $\sim (\hat{p}\cdot \hat{z})\xi/\dS$ times, where $\hat{z}$ denotes the normal to the interface and $\hat{p}$ the direction of the electronic momentum. Modes which propagate mostly parallel to the interface ($\hat{p}\cdot \hat{z}\sim 0$) are little affected by the exchange coupling of the ferromagnet. This effectively smoothens the vanishing of the superconducting gap with increasing exchange coupling and results in a second-order transition.

A qualitative understanding of our results can be obtained from semiclassical estimates of the proximity-induced effective Zeeman field $B_{\mathrm{eff}}$ and induced superconducting gap $\Delta_\mathrm{ind}$. For these estimates, we assume a N-SC interface with unit transparency. Even in the absence of an interface potential, the transparency of the interface depends on the velocity mismatch between the N and SC layers. Unit transparency is only found for equal Fermi velocities on both sides. For typical nanowire materials, the Fermi velocities $\vN$ and $\vS$ of the two layers can indeed be similar in magnitude, $\vN\approx\vS$, despite the large difference in Fermi wavevectors reflecting the widely different electron densities in the semiconductor and the superconductor. The reason is that this difference in Fermi wavevectors is offset by a comparable difference in effective masses. Thus assuming unit transparency of the N-SC interface, the superconducting gap $\Delta_\mathrm{ind}$ induced in the normal region is proportional to the fraction of time a mode spends in the superconductor  \cite{Kiendl2019},
\begin{equation}
\Delta_\mathrm{ind} = \frac{\tS}{\tS+\tN} \Delta . \label{eq:Deltaind}
\end{equation}
Here $\tau_i = d_i/v_{i_z}$ with $i=$N, S. Notice that the induced superconducting gap depends on the $z$ component $v_{i_z}$ of the velocity and is thus specific for each mode in the normal layer. (Due to the large mismatch in densities, typical modes correspond to almost normal trajectories in the superconductor, so that $v_{S_z}\approx \vS$ \cite{Kiendl2019}.)

\begin{figure}[b]
\centering
\includegraphics[width=0.75\columnwidth]{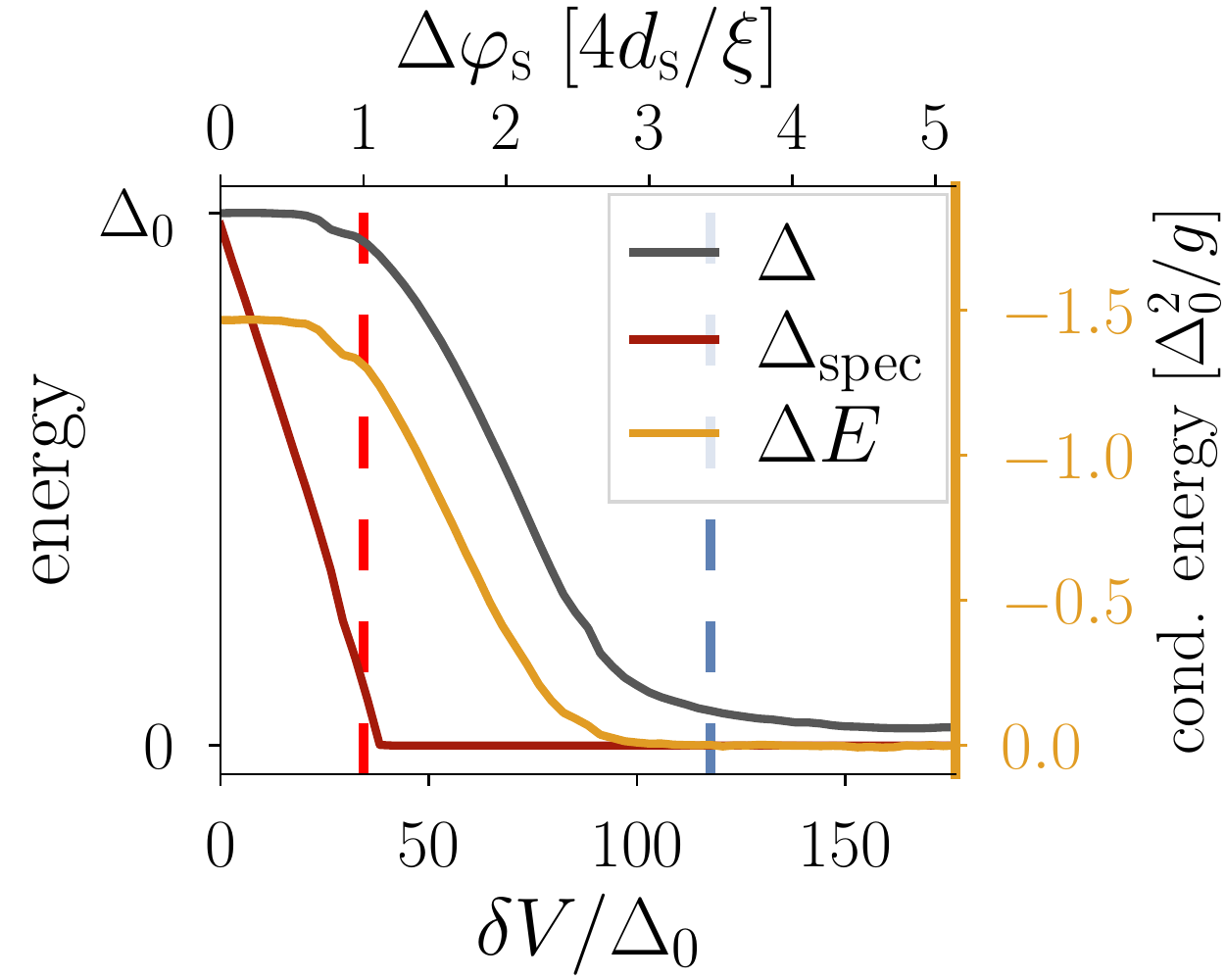}
\caption{Self-consistent approach to superconductivity in a SC-F nanowire, including the self-consistent pairing correlations ${\Delta}$ and the spectral gap $\Delta_{\mathrm{spec}}$ as a function of the scattering phase difference $\Delta\phiS$. $\Delta_{\mathrm{spec}}$ vanishes close to $\Delta\phiS\sim 4\dS/\xi$, while the pairing correlations ${\Delta}$ remain finite. The condensation energy $\Delta E$ goes to zero before the pairing correlations ${\Delta}$ vanish, corresponding to a weakly first-order phase transition. Calculations were done for $\kS\dS/\pi = 27.13$ and an F layer of infinite thickness. For other parameters, see Table \ref{tab:real_parameters}.	\label{fig:NSF_gap_curve}}
\end{figure}

Similarly, the effective Zeeman field induced by the ferromagnetic insulator can be obtained from Bohr-Sommerfeld quantization as
\begin{equation}
B_{\mathrm{eff}} = \frac{\hbar\Delta\varphi}{4(\tS+\tN)}.
\label{eq:Beff}
\end{equation}
where $\Delta\varphi = \varphi_\uparrow - \varphi_\downarrow$ is the difference between the spin-dependent scattering phases $\varphi_\sigma$. In accordance with the discussion above and as for the induced superconducting gap, this effective Zeeman field is mode dependent.

\begin{figure*}[t]
\centering
\includegraphics[width=\textwidth]{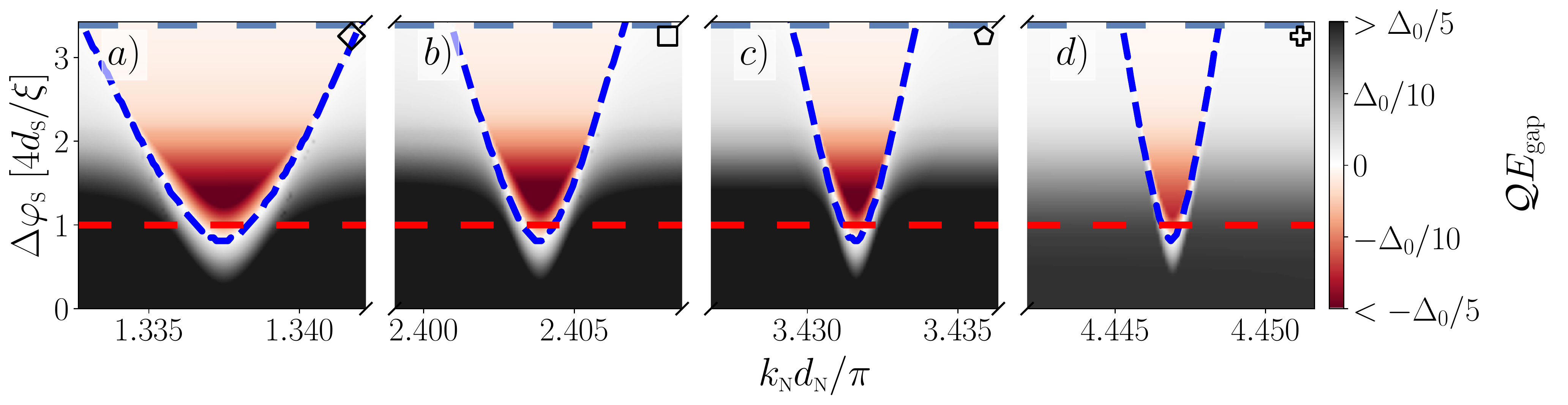}
\caption{Phase diagram for the {N-SC-F} arrangement. The magnitude of the overall gap $E_\mathrm{gap}$, multiplied by the topological invariant $\mathcal{Q}= \pm 1$, is color coded as a function of the phase difference $\Delta \phiS$ and the number $\kN\dN/\pi$ of occupied transverse modes in the normal layer. The four panels highlight the parameter ranges where the first four transverse modes in N become populated. At the optimal chemical potentials, the onset of the topological phase takes place close to the semiclassical estimate $\Delta\phiS = 4 \dS/\xi$, as indicated by the red dashed line. The gray dashed line labels the magnitude of $\Delta\phiS$, where the superconducting condensation energy changes sign and superconductivity is fully suppressed by the adjacent ferromagnetic insulator. Phase transitions are denoted by a blue dashed line. The symbol in the top right corner of the panels indicates the parameter choice as shown in Fig.\ \ref{fig:resonance_all}. The number of transverse modes in S was fixed at $\kS\dS/\pi = 27.16$ for all panels. For other parameters, see Table \ref{tab:real_parameters}.
\label{fig:NSF_phase_diagram}}
\end{figure*}

We are now in a position to discuss the emergence of topological superconductivity in the three geometries shown in Fig.\ \ref{fig:hexagonal_wire}. The phase diagram for the SC-N-F geometry, obtained from our detailed theory described in Sec.\ \ref{sec:model}, is shown in Fig.\ \ref{fig:SNF_phase_diagram}. In this geometry, the superconducting and ferromagnetic layers are spatially separated. This minimizes the detrimental effect of the spin splitting induced by the ferromagnetic insulator on the superconductor. As a result, we can deduce the phase difference required for topological superconductivity directly from Eqs.\ (\ref{eq:Deltaind}) and (\ref{eq:Beff}), with $\Delta$ equal to the unperturbed superconducting pairing $\Delta_0$ of the superconductor. For optimal chemical potential, modes are expected to become topological when $B_{\mathrm{eff}}\geq\Delta_\mathrm{ind}$ \cite{Lutchyn2010,Oreg2010}. Thus, using Eqs.\ (\ref{eq:Deltaind}) and (\ref{eq:Beff}), the condition for topological superconductivity becomes
\begin{equation}
\Delta\varphi \geq  \frac{4\Delta \dS}{\hbar \vSz} \approx  \frac{4\dS}{\xi}. \label{eq:phasediff_estimate}
\end{equation}
In the last step, we used $\vSz\approx\vS$ due to the large difference in Fermi wavevectors between semiconductor and superconductor. We find that this minimal phase difference $\Delta\varphi$ is in good agreement with the phase diagram in Fig.\ \ref{fig:SNF_phase_diagram}, see the red dashed line in panel (a).  Away from the optimal chemical potential, the effective Zeeman splitting required to induce a topological superconducting phase increases, cp.\ \cite{Lutchyn2010,Oreg2010}, qualitatively explaining the shape of the topological regions. The four panels detail the parameter ranges of the phase diagram, where the first four transverse modes of the N layer become populated with increasing $\kN$. In agreement with expectations, it is these regions where topological superconductivity emerges. We find that the induced topological gap becomes smaller for higher transverse modes. This reflects variations in the ratio of the Fermi velocities in the semiconductor and the superconductor.  

For the other layer stackings, the direct proximity of ferromagnetic insulator and superconductor suppresses superconductivity, eventually driving the superconducting layer normal. According to Eq.\ (\ref{eq:phasediff_estimate}), the scattering phase difference necessary to overcome the superconducting gap of a mode with velocity $\vSz \approx \vS \cos \phi $ is 
\begin{equation}
\Delta\varphi =  \frac{4\dS}{\xi\cos\phi}, \label{eq:phasediff_selfcon}
\end{equation}
where $\cos \phi = [{1-(k_x/\kS)^2}]^{1/2}$. Modes with a higher longitudinal momentum $k_x$ impinge less often on the interface with the ferromagnet, thus requiring larger exchange couplings to overcome their superconducting gap. With increasing spin splitting in the ferromagnet, the superconducting gaps of the modes successively close, which in turn affects the overall pairing correlations. For these stackings, we thus determine the pairing strength $\Delta$ from a self-consistent treatment. Once the self-consistent pairing strength is determined, we can again apply Eqs.\ (\ref{eq:Deltaind}) and (\ref{eq:Beff}) to estimate the minimal phase difference for entering the topological phase. 

The result of such a self-consistent calculation for the superconducting gap is shown in Fig.\ \ref{fig:NSF_gap_curve} for a ferromagnet of thickness $d_F \to \infty$. We find that the spectral gap closes with increasing spin-dependent phase difference. This occurs when $\Delta\varphi \approx  {4\dS}/{\xi}$, where the coherence length is computed with the bare superconducting pairing strength $\Delta_0$ in the absence of the ferromagnetic layer. This is consistent with the fact that for the value of $\Delta\varphi$ at which the spectral gap closes, the self-consistent pairing strength $\Delta$ is only weakly suppressed. The self-consistent pairing strength persists to stronger exchange fields exerted by the ferromagnet. We find that the condensation energy $\Delta E$ drops to zero prior to a complete suppression of the self-consistent pairing strength, predicting a phase transition into the normal state that is weakly first order. (For a uniform exchange field, the transition into the normal state is first order and takes place before the spectral gap closes \cite{Clogston1962,Chandrasekhar1962}. The situation for a superconductor coupled to a ferromagnetic insulator is different \cite{Tokuyasu1988} and may reverse the order in which the spectral gap closes and the order parameter vanishes.)

\begin{figure*}[t]
\centering
\includegraphics[width=\textwidth]{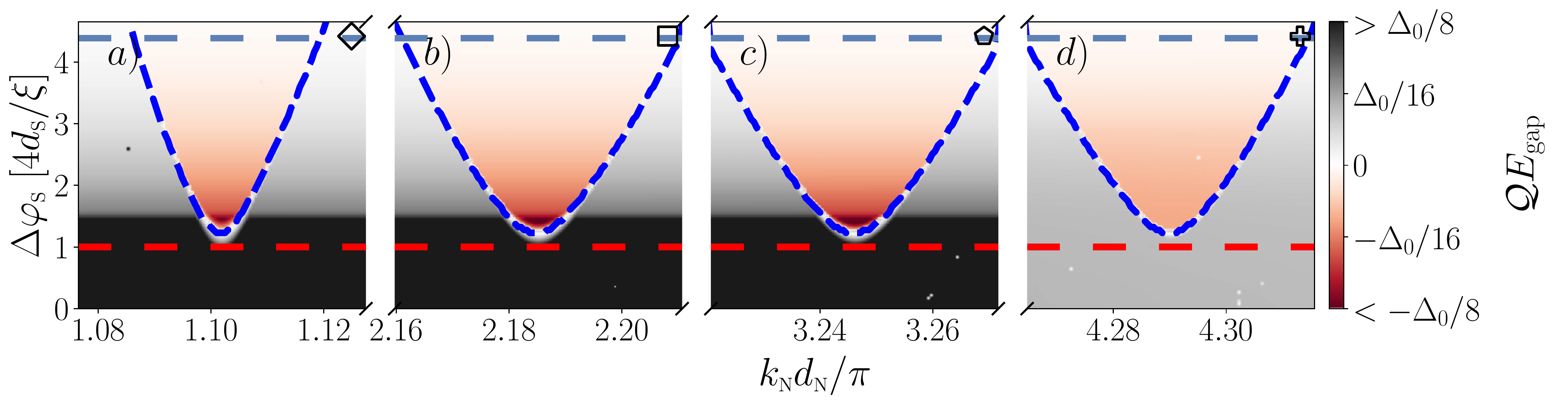}
\caption{Phase diagram for the { N-F-SC} arrangement. The magnitude of the overall gap $E_\mathrm{gap}$, multiplied by the topological invariant $\mathcal{Q}= \pm 1$, is color coded as a function of the phase difference $\Delta \phiS$ and the number $\kN\dN/\pi$ of occupied transverse modes in the normal layer. The four panels highlight the parameter ranges where the first four transverse modes in N become populated. At the optimal chemical potentials, the onset of the topological phase takes place close to the semiclassical estimate $\Delta\phiS = 4 \dS/\xi$, as indicated by the red dashed line. The gray dashed line labels the magnitude of $\Delta\phiS$, where the superconducting condensation energy changes sign and superconductivity is fully suppressed by the adjacent ferromagnetic insulator. Due to the finite thickness of F, this occurs at a higher $\Delta\phiS$ value than in Fig.\ \ref{fig:NSF_phase_diagram}. Phase transitions are denoted by a blue dashed line. The symbol in the top right corner of the panels indicates the parameter choice as shown in Fig.\ \ref{fig:resonance_all}. In all panels, the number of transverse modes in S was fixed at $\kS\dS/\pi = 27.61$ and the thickness of F was chosen to satisfy $\kF\dF=1.22$. For other parameters, see Table \ref{tab:real_parameters}. \label{fig:NFS_phase_diagram}}
\end{figure*}

The resulting phase diagram for N-SC-F stacking is shown in Fig.\ \ref{fig:NSF_phase_diagram}. Also in this case, we find regions of topological superconductivity whenever a new transverse mode opens in the semiconductor. However, the topological regions not only have a smaller gap than in the SC-N-F stacking, but are also limited to a much smaller parameter range. This limitation is imposed by the small values of $\Delta\varphi$ that are compatible with substantial superconducting correlations. It is also interesting to compare this result to a recent result \cite{Poyhonen2020} that topological superconductivity requires Zeeman fields that locally exceed the pairing strength of the superconductor. This result precludes topological superconductivity in a bipartite N-SC structure, which applies the Zeeman splitting to the superconductor only. A N-SC-F structure can still support topological superconductivity since the ferromagnetic proximity effect is not identical to the application of a uniform Zeeman field to the superconductor. 

Finally, the semiclassical considerations and estimates performed in this section do not apply directly to the N-F-SC arrangement. In this stacking, the ferromagnetic insulator effectively acts as a potential barrier separating the N and SC layers. The superconductor affects the semiconductor only when the ferromagnet is sufficiently thin, satisfying $\kF\dF \lesssim 1$, with $\kF$ the wavevector characterizing the wave-function decay in F. In this case, the system can enter a topological superconducting phase, albeit with a smaller gap than for the other two arrangements due to the reduced mixing of superconductivity and spin-orbit coupling. This arrangement is limited to small scattering phase differences for the same reason as for the N-SC-F stacking, as the superconductor and the ferromagnet again share an interface. Numerically, the limiting value of $\Delta\varphi$ is somewhat larger than in the N-SC-F arrangement. This difference is a consequence of the small thickness $\dF$ of the F region in the N-F-SC arrangement, which reduces the detrimental effect of F on SC. 

\section{Model and calculations} \label{sec:model}

In our detailed calculations, we model nanowires, which are infinitely extended in the $x$-direction and composed of three layers stacked along the $z$-direction: a semiconductor (N), a superconductor (SC) and an insulating ferromagnet (F). The extent in the $y$-direction is assumed small enough for a single mode to be occupied. In the conventional Nambu basis $\Psi=(\psi_{\uparrow} , \psi_{\downarrow} , \psi_{\downarrow}^{\dagger} , -\psi_{\uparrow}^{\dagger})^T$, the BdG Hamiltonian takes the form
\begin{equation}
\mathcal{H} = H_0 (z) \tau_z + \Delta (z) \tau_x + \alpha(z) k_x \sigma_x \tau_z,
\end{equation}
where $\boldsymbol{\tau}$ and $\boldsymbol{\sigma}$ are Pauli matrices acting in particle-hole and spin space, respectively. The superconducting pairing $\Delta (z)$ and the strength $\alpha(z)$ of the Rashba spin-orbit coupling are assumed piecewise constant and nonzero only within their respective nanowire layers,
\begin{equation}
\Delta(z)=\begin{cases}
\Delta_0 & z\in \mathrm{SC}\\
0 & \textrm{else,}
\end{cases}
\end{equation}
and 
\begin{equation}
\alpha(z)=\begin{cases}
\alpha & z\in \mathrm{N}\\
0 & \textrm{else.}
\end{cases}
\end{equation}
Finally, $H_0(z)$ is given by
\begin{equation}
H_0(z) =\! \sum_{i=x,y,z} p_i \frac{1}{2 m^*(z)} p_i + V_0 (z),
\end{equation}
with $p_i$ the momentum along the $i$-direction, $m^*(z)$ the effective mass, and  $V_0(z)$ the band offset. For simplicity, we assume equal effective masses for the SC and the F layers,
\begin{equation}
m^*(z)=\begin{cases}
\mN & z\in \mathrm{N}\\
\mS & z\in \textrm{SC, F.}
\end{cases}
\end{equation}
The band offset is expressed via the corresponding Fermi wavevectors $\kN$ and $\kS$ in N and SC, respectively, and via the inverse decay length $\kF$ in F, 
\begin{equation}
V_{0}(z)  =\begin{cases}
\;\;\; -\frac{\kN^{2}}{2\mN} & z\in \mathrm{N}\\[0.1cm]
\;\;\; -\frac{\kS^{2}}{2\mS} & z\in \mathrm{SC}\\[0.1cm]
\frac{\kF^{2}}{2\mS} + \delta V\sigma_z     & z \in \mathrm{F},
\end{cases}
\end{equation}
where we have set $\hbar = 1$. Thus, the ferromagnetic insulator F is modeled as a spin-dependent potential barrier characterized by the spin splitting $\delta V$. Previous studies point out that the induced exchange coupling in a superconductor cannot be fully explained in terms of the large optical band gap of the ferromagnetic insulator and may also involve direct coupling of the electrons with atomic exchange fields \cite{Tokuyasu1988}. Still, we model the ferromagnet as a spin-dependent potential barrier and use the spin splitting $\delta V$ as a phenomenological parameter to parametrize the ferromagnetic proximity effect.

The thicknesses of the layers are denoted by $\dN$, $\dS$, and $\dF$. We take $\dF$ to be infinite in the N-SC-F and SC-N-F arrangements, which is appropriate as long as the thickness of the ferromagnet is large compared to the penetration depth into F. For the N-F-SC stacking, we assume a small and finite $\dF$ to allow for coupling between the N and SC regions. Finally, we note that the number of occupied transverse modes in the N and SC layers can be estimated as $\kN \dN / \pi$ and $\kS \dS / \pi$, respectively.

We summarize the parameters used for our numerical calculations in Table \ref{tab:real_parameters}. These parameters are chosen to resemble parameters in InAs, Al ($\Delta_0=0.34$meV), and EuS.

\begin{table}[t]
\centering
\begin{tabular}{
>{\centering\arraybackslash}m{0.8cm}|
>{\centering\arraybackslash}m{2.6cm}||
>{\centering\arraybackslash}m{0.8cm}|
>{\centering\arraybackslash}m{1.7cm} m{-1.5cm}}
$\mN$ & $0.1\,m_{e}$  &  $\alpha$ & $0.035 \, \Delta_0 \, \xi$ &
\\[0.2cm]
\hline & & & 
\\[-0.3cm]
$\mS$ & $1.16\,m_{e}$  & $l_\rmsss{SOC}$ & $\;4.7\cdot 10^{-3}\xi$ &  
\\[0.2cm]
\hline & & & 
\\[-0.3cm]
$\kN$ & $\; (0 \sim 5.8)\cdot 10^3 \,\xi^{-1} \;$  & $\dN$ & $\;2.4\cdot 10^{-3}\xi$ &
\\[0.2cm]
\hline & & & 
\\[-0.3cm]
$\kS$ & $ \sim 7\cdot 10^4 \, \xi^{-1}$  & $\dS$ & $\; 1.2\cdot 10^{-3}\xi\;$ &
\\[0.2cm]
\hline & & & 
\\[-0.3cm]
$\kF$ & $2.6 \cdot 10^4 \, \xi^{-1}$  & $\dF$ & $\;4.7\cdot 10^{-5}\xi$ &
\\[0.2cm]
\hline & & & 
\\[-0.3cm]
$\Delta_0$ & $0.34$meV  & $\xi$ & $\;3680$nm &  
\end{tabular}
\caption{Parameters used in the numerical calculations.\label{tab:real_parameters}}
\end{table}

\subsection{Transverse modes} \label{sec:transversal_modes}

The eigenfunctions in the longitudinal direction are plane waves for all three layers. The eigenfunctions in the transverse direction (transverse modes) can thus be obtained by matching piecewise solutions in the three layers. For the N layer, the eigenfunctions separate into the particle and hole sectors. The electron wave function is
\begin{align}
\vect{\psi}_{\rmsss{N},e}(z,\varepsilon) =\,\, & \vu{e}_{+}\left(c_{e}^{+}e^{i\kNz^{+}\!(\varepsilon)z}+c_{e}^{+'} e^{-i\kNz^{+}\!(\varepsilon)z}\right) \nonumber \\
&+ \vu{e}_{-}\left(c_{e}^{-}e^{i\kNz^{-}\!(\varepsilon)z}+c_{e}^{-'}e^{-i\kNz^{-}\!(\varepsilon)z}\right),
\end{align}
while the hole wave function takes the form
\begin{align}
\vect{\psi}_{\rmsss{N},h}(z,\varepsilon)=\,\, & \vu{e}_{+}\left(c_{h}^{+}e^{i\kNz^{+}\!(-\varepsilon)^{*}z}+c_{h}^{+'}e^{-i\kNz^{+}\!(-\varepsilon)^{*}z}\right) \nonumber \\
 &+ \vu{e}_{-}\left(c_{h}^{-}e^{i\kNz^{-}\!(-\varepsilon)^{*}z}+c_{h}^{-'}e^{-i\kNz^{-}\!(-\varepsilon)^{*}z}\right).
\end{align}
with $\vu{e}_{\pm}=\frac{1}{\sqrt{2}}(1 ,\pm 1)^T$ the eigenspinors of $\sigma_x$. The overall eigenfunction is
\begin{equation}
\vect{\psi}_{\rmsss{N}}(z,\varepsilon) = \vu{e}_{e} \otimes \vect{\psi}_{\rmsss{N}, e}(\mathbf{r},\varepsilon) + \vu{e}_{h} \otimes \vect{\psi}_{\rmsss{N}, h}(\mathbf{r},\varepsilon),
\end{equation}
where $\vu{e}_{e}=(1,0)^T$ and $\vu{e}_{h}=(0,1)^T$ are spinors in particle-hole space. In the SC, the Hamiltonian splits into two blocks that can be labeled by their $\sigma_z$ eigenvalue $\nu=\pm1$,
\begin{align}
\vect{\psi}_{\rmsss{S}, \nu}(\mathbf{r},\varepsilon)=\,\,&\hat{\bm{\varepsilon}} \left(d_{\nu}^{+}e^{i\kSz^{+}\!(\varepsilon)z}+d_{\nu}^{+'}e^{-i\kSz^{+}\!(\varepsilon)z}\right) \nonumber\\
&+\left(\tau_{x}\hat{\bm{\varepsilon}} \right )\left(d_{\nu}^{-}e^{i\kSz^{-}\!(\varepsilon)z}+d_{\nu}^{-'}e^{-i\kSz^{-}\!(\varepsilon)z}\right),
\end{align}
with $\hat{\bm{\varepsilon}} = (e^{i\beta} , 1)^T$ the eigenspinors of the $2\times 2$ superconducting BdG Hamiltonian and $\beta = \arccos (\varepsilon/\Delta)$. The overall SC eigenfunction then takes the form
\begin{equation}
\vect{\psi}_{\rmsss{S}}(\mathbf{r},\varepsilon) = \vect{\psi}_{\rmsss{S}, +}(\mathbf{r},\varepsilon) \otimes \vu{e}_{\uparrow}  + \vect{\psi}_{\rmsss{S}, -}(\mathbf{r},\varepsilon) \otimes  \vu{e}_{\downarrow} .
\end{equation}
The spinors $\vu{e}_{\uparrow/\downarrow}$ are $(1 , 0)^T$ and $(0,1)^T$ in spin space, respectively. Note that $\vu{e}_{\uparrow/\downarrow}$ correspond to $\uparrow/\downarrow$ electron in the particle sector and to $\downarrow/\uparrow$ holes in the hole sector.
Finally, the wave function in F does not mix particles and holes or spin projections $\sigma_z$,
\begin{equation}
\vect{\psi}_{\rmsss{F}}(\mathbf{r},\varepsilon)\! =\!\!\!\! \sum_{\substack{\tau=e,h \\ \sigma=\uparrow/\downarrow}}\!\!  \vu{e}_{\tau} \otimes \vu{e}_{\sigma} \! \left (\! f_{\tau, \sigma} e^{\kappa_{\sigma_z}(\tau \varepsilon)z} \! + \!  f'_{\tau, \sigma} e^{-\kappa_{\sigma_z}(\tau \varepsilon)z}\! \right )\!,
\end{equation}
where $\tau=e,h$ acts as $\pm1$ when not an index. The transverse momenta $\kNz^{\pm}$, $\kSz^{\pm}$ and $\kappa_{\sigma_z}$ are defined as
\begin{align} 
\kNz^{\pm}(\varepsilon) & =\sqrt{\kN^2-k_{x}^{2}\mp 2\alpha m_{\rmsss{N}} k_{x}-k_{y}^{2}+2m_{\rmsss{N}}\varepsilon}, \\
\kSz^{\pm}(\varepsilon) & =\sqrt{\kS^2-k_{x}^{2}-k_{y}^{2}\pm 2m_{\rmsss{S}}\sqrt{\varepsilon^{2}-\Delta^{2}}}, \\
\kappa_{\uparrow/\downarrow}(\varepsilon) & =\sqrt{\kappa^2_\rmsss{F}+k_{x}^{2}+k_{y}^{2}-2 m_{\rmsss{S}} \left(\varepsilon \pm \delta V\right)}. \label{eq:decayF}
\end{align}
Throughout the paper, we set $k_y=0$, effectively rendering the nanowire two-dimensional.
Extending the calculation to finite $k_y$ would be straightforward and amounts to a redefinition of $\kS$, $\kN$, and $\kF$. Even in such a three-dimensional calculation, low-$k_y$ modes are expected to exhibit an enhanced mixing of superconductivity, spin-orbit coupling, and spin-splitting as they scatter more strongly between the three nanowire constituents.

\subsection{Wave-function matching}

We find the eigenfunctions and eigenenergies of the nanowire by matching wave functions and ensuring current conservation at the interfaces. One of the $3 \cdot 8$ free coefficients of the wave function sets the overall prefactor which is ultimately fixed by the normalization condition. The remaining coefficients as well as the energy $\varepsilon$ (for fixed $k_x$) are determined by (i) the $2\cdot2\cdot 4$ equations accounting for continuity of the wave functions and current conservation at the two internal interfaces, and (ii) the vanishing of the wave function at the outer interfaces adds $2 \cdot 4$ equations. 

\subsection{Ferromagnetic proximity effect}

We quantify the strength of the induced exchange coupling by the difference in the scattering phases from the interface with the ferromagnet for the two spin directions. To this end, consider an interface between a half-infinite ferromagnetic insulator and a normal region without spin-orbit coupling or superconductivity. Matching the wavefuncions at the interface yields a spin- and momentum-dependent scattering phase of
\begin{equation}
\varphi_{\beta,\sigma}(k_x) = \pi - 2\arctan\left( v_{\beta_z} / v_{\rmsss{F_z},\sigma} \right), \label{eq:scattering_phase}
\end{equation}
where $v_{\beta_z} = k_{\beta_z} / m_{\beta}$ with $\beta=\mathrm{N},\mathrm{S}$ and $v_{\rmsss{F_z},\sigma} = \kappa_{\sigma_z} / \mS$ and with the momenta evaluated at $\alpha=0$, $\varepsilon=0$, and $\Delta=0$. The scattering phase difference is then
\begin{equation}
\Delta \varphi_{\beta}(k_x) = \varphi_{\beta,\uparrow}(k_x) - \varphi_{\beta,\downarrow}(k_x) . \label{eq:scattering_phasediff}
\end{equation}
In the N-SC-F and N-F-SC arrangements, where SC and F share an interface, we use $\Delta \phiS=|\Delta \phiS(0)|$ to quantify the induced exchange coupling. For the SC-N-F arrangement, where F only shares an interface with N, we use $\Delta \phiN=|\Delta \phiN(0)|$. 

Within our model, $\Delta \phiN$ and $\Delta \phiS$ are determined by the Fermi velocities for the three types of layers. As mentioned above, we take these to be comparable for the three types of layers. We note, however, that this may overestimate the correlations between $\Delta \phiN$ and $\Delta \phiS$  in the experimental samples. In addition to the ferromagnetic proximity effect involving atomic exchange fields, band bending effects not included in our modeling may modify the N-F and SC-F interfaces differently.  

\subsection{Self-consistent treatment of superconductivity} \label{sec:sc_sc}

In the SC-N-F arrangement, there is no direct interface between the SC and F layers. Since most modes in the SC decay rapidly into N as a result of the large disparity in electron densities, the ferromagnet affects the superconducting pairing correlations only weakly. In contrast, effects of self-consistency become important for the N-SC-F and N-F-SC arrangements. For these geometries, the presence of N has only a small influence on the strength of superconducting correlations. For this reason, we consider the interplay between superconductivity and magnetism within a reduced model of a bipartite SC-F nanowire. In Fig.\ \ref{fig:NSF_gap_curve}, we present results for an F layer, which is infinitely extended in the $z$ direction. Due to the rapid decay of the mode wave functions into the ferromagnetic insulator, this is an accurate model for the N-SC-F geometry, and provides an upper bound on the suppression of superconductivity by the ferromagnet for the  N-F-SC arrangement. (We have also considered finite F slabs and find that the behavior is similar up to changes in the spin-dependent phase difference by factors of order unity.) 

We obtain the self-consistent order parameter $\Delta$ of the SC-F structure from the equation  
\begin{equation}
\Delta(\mathbf{r},\delta V)=g\expval{\psi_{\uparrow}(\mathbf{r}) \psi_{\downarrow}(\mathbf{r})}_{\delta V}, \label{eq:delta_of_r}
\end{equation}
with $g<0$ denoting the strength of the attractive interaction and making the dependence on the spin splitting $\delta V$ in F explicit. Using the modes obtained in Sec.\ \ref{sec:model}, we iteratively solve for $\Delta(\mathbf{r},\delta V)$ starting with an initial superconducting pairing $\Delta_0$. As a result of the strictly local interaction $g$, Eq.\ (\ref{eq:delta_of_r}) will yield a spatially oscilllatory $\Delta(\mathbf{r},\delta V)$. The physically relevant pairing strength $\Delta(\delta V)$ is obtained by averaging over these oscillations and it is this average ${\Delta}$ which is used as input in subsequent iteration steps. The value of $g$ is chosen such that ${\Delta}(0)=\Delta_0$.  Convergence is attained once the difference between the input ${\Delta}(\delta V)$ and the corresponding output falls below a threshold.

\begin{figure*}[t]
\centering
\includegraphics[width=\textwidth]{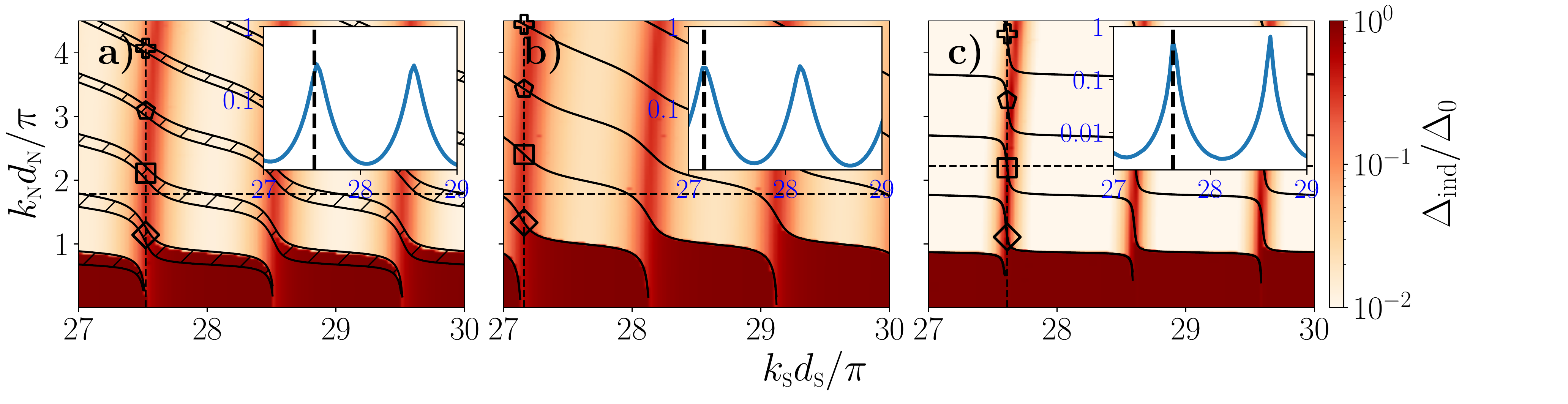}
\caption{Induced gap $\Delta_{\mathrm{ind}}$ in the absence of spin splitting as a function of the mode occupations $\kS \dS/\pi$ and $\kN \dN/\pi$ in the N and SC layers, respectively. The insets show linecuts along the dashed horizontal lines, with vertical dashed lines corresponding between the main panels and the insets. Note that the position of the resonance peak in $\kS \dS/\pi$ shifts slightly with $\kN \dN/\pi$. {(a)} {SC-N-F} arrangement: Resonance at $\kN \dN/\pi = 1.78$ and $\kS \dS/\pi = 27.52$.
{(b)} {N-SC-F} arrangement: Resonance at $\kN \dN/\pi = 1.78$ and $\kS \dS/\pi = 27.16$. {(c)} {SC-F-N} arrangement: Resonance at $\kN \dN/\pi = 2.23$ and $\kS \dS/\pi = 27.61$. The symbols marking the topological regions indicate the choice of parameter values in the panels in Figs.\ \ref{fig:SNF_phase_diagram}, \ref{fig:NSF_phase_diagram}, and \ref{fig:NFS_phase_diagram}, which are labeled by corresponding symbols. For other parameters, see Table \ref{tab:real_parameters}.
\label{fig:resonance_all}}
\end{figure*}

The resulting self-consistent solutions for ${\Delta}(\delta V)$ are presented in Fig.\ \ref{fig:NSF_gap_curve}, showing a suppression of the gap function as the spin-splitting in F increases. The spectral gap $\Delta_{\mathrm{spec}}$ is suppressed even more rapidly and vanishes when ${\Delta}(\delta V)$ is still finite. Modes with a larger transverse momentum (and hence smaller $k_x$) frequently scatter off the ferromagnet, rapidly suppressing the excitation gap in these modes. It is thus the effect of the spin splitting on these modes, which causes the spectral gap $\Delta_{\mathrm{spec}}$ to vanish. More quantitatively, we find that $\Delta_\mathrm{spec}$ vanishes close to the estimated value of $\Delta\phiS =4 {\dS}/{\xi}$ of the scattering phase difference (with $\xi$ computed with $\Delta_0$). In contrast, modes with higher $k_x$ scatter less often off the ferromagnet and effectively sustain nonzero pairing correlations. The intermediate regime of zero $\Delta_{\mathrm{spec}}$ and finite ${\Delta}(\delta V)$ corresponds to a gapless superconductor. Here, inclusion of spin-orbit coupling may reopen the gap and drive the system into a topological phase. 

Eventually, as the spin-splitting increases further, ${\Delta}(\delta V)$ decreases to zero. We also calculate the condensation energy of the superconductor as the difference in the ground state energies of the system with and without superconductivity. We find that in our geometry, the condensation energy vanishes when the superconducting pairing is already strongly suppressed, but not yet zero. This indicates a weakly first-order phase transition to the normal state \cite{Hao1991}, with a finite, but small discontinuity in the order parameter. This weakly first-order phase transition is a consequence of the finite width $d_S$ and the associated discrete mode structure in the superconductor. The transition is second order if the discreteness of the transverse modes in SC can be neglected.

\section{Phase Diagrams} \label{sec:Phase_Diags}

Performing the numerical calculations outlined in the previous section, we extract topological phase diagrams for the three layer arrangements. We first consider the induced gap $\Delta_{\mathrm{ind}}$ of the system as a function of the number of propagating modes $\kN\dN/\pi$ and $\kS\dS/\pi$ in the normal and the superconducting regions. ($\Delta_{\mathrm{ind}}$ is computed in the absence of any spin splitting $\delta V$.) Figure \ref{fig:resonance_all} shows corresponding color-scale plots of  $\Delta_{\mathrm{ind}}$ for the three layer arrangements. $\Delta_{\mathrm{ind}}$ is found to be only weakly dependent on $\kN$, except when the first mode in N begins to be occupied. In contrast, there is an oscillatory dependence on $\kS$, which originates from resonances between the transverse modes in N and SC \cite{Reeg2017,Reeg2018,Kiendl2019}. This oscillatory dependence is emphasized by the insets in Fig.\ \ref{fig:resonance_all}, which display line cuts of $\Delta_{\mathrm{ind}}$ as a function of $\kS\dS/\pi$. 

The topological nature of the gap can be extracted by computing the topological invariant $\mathcal{Q}=\mathrm{sgn}(\mathrm{Pf}\left(\mathcal{H}\right))$, which corresponds to the fermion parity of the ground state of the system \cite{Moore1991,Read2000,Kitaev2001}. Within our continuum model, there is only one time-reversal invariant point $k_x=0$, where this change of fermion parity can occur. The hatched regions superimposed on the color-scale plots in Fig.\ \ref{fig:resonance_all} indicate the extent of the topological superconducting regions in the presence of a nonzero spin-dependent phase difference (with $\Delta\varphi$ equal to the largest value included in Figs.\ \ref{fig:SNF_phase_diagram}, \ref{fig:NSF_phase_diagram}, and \ref{fig:NFS_phase_diagram}.) The region has a visible extent only for the SC-N-F arrangement, while it is too small to be visible for the other two layer arrangements. The phase transition lines are only weakly dependent on $\kS\dS/\pi$ away from the resonances and change by one unit in $\kN\dN/\pi$ around the resonances. This indicates that a new mode becomes populated in the superconductor at these resonances. This behavior appears most clearly in the N-F-SC arrangement, where N and SC are separated by the potential barrier of F.

To display the dependence on the strength of the ferromagnetic proximity effect, we fix the number of occupied modes in the superconducting layer to one of the resonances and plot the spectral gap, multiplied by the topological invariant, as a function of both the mode number in the semiconducting layer and the spin-dependent phase difference as shown in Figs.\ \ref{fig:SNF_phase_diagram}, \ref{fig:NSF_phase_diagram}, and \ref{fig:NFS_phase_diagram} and discussed in Sec.\ \ref{sec:phys_disc} above. The four panels in these phase diagrams are labeled by symbols, which indicate the corresponding parameter choice in Fig.\ \ref{fig:resonance_all}. 

\section{Conclusion}
\label{sec:conclusion}

Tripartite nanowires proximity coupling a semiconducting core to epitaxial superconductor and ferromagnetic-insulator layers may obviate the need for applying an external magnetic field for realizing topological superconductivity and thereby open new design opportunities for Majorana-based devices. At the same time, the additional epitaxial ferromagnetic insulator adds new material-science challenges. Our study aimed at understanding the observation of zero-bias peaks in a recent experiment \cite{Vaitieknas2020} on such tripartite nanowires, which may constitute evidence for topological superconductivity.

As shown schematically in Fig.\ \ref{fig:hexagonal_wire}, the relevant region of the nanowire includes all three possible interfaces between the three constituents, N, SC, and F. To elucidate and differentiate the mechanisms by which topological superconductivity can emerge in this structure, we focus on the three paradigmatic stackings SC-N-F, N-SC-F, and N-F-SC (see also \cite{Maiani2020}). Our approach treats these stackings within a microscopic wave-function approach, but neglects band bending effects (which, however, would be the underlying reason why the intersection region of the three constituents is most relevant for the emergence of topological superconductivity \cite{Woods2020a,Escribano2020,Liu2020}). We also focus on clean structures in the absence of bulk or interface disorder. 

We find that the SC-N-F arrangement has by far the largest topological superconducting region in parameter space. In this arrangement, superconducting correlations and spin splitting are induced in N by proximity from the SC and F layers, respectively. At the same time, the intermediate N region effectively shields the SC from the detrimental influence of the ferromagnetic insulator, as most modes in the SC are only weakly coupled to the N region due to the much larger electron density in SC than in N. In this arrangement, topological superconductivity therefore emerges by essentially the same mechanism as previously considered for hybrid semiconductor nanowires \cite{Lutchyn2010,Oreg2010}, with spin splitting now emerging by proximity rather than by applied Zeeman field. This mechanism is also expected to be rather robust against modifications of the model.  In particular, we expect disorder in the superconductor to increase the strength and the robustness of the superconducting proximity effect \cite{Kiendl2019}. 

The two other stackings involve direct SC-F couplings, requiring a self-consistent treatment of  superconducting correlations. The ferromagnet insulator has a strongly detrimental effect on the superconductor, thereby severely limiting the spin splitting that can be induced in N without suppressing superconducting correlations entirely. We still find that topological superconductivity can emerge for both, the N-SC-F and the N-F-SC stackings, but only in very limited regions of parameter space. 

For the N-SC-F stacking, disorder in the SC layer may further limit the topological region in parameter space. While the ferromagnetic proximity effect differentiates between modes of the SC in the clean limit, this is no longer the case when the SC layer is disordered. The ferromagnetic proximity effect will then be essentially equivalent to the application of a Zeeman field applied to the SC layer, a situation for which a recent theorem \cite{Poyhonen2020} precludes topological superconductivity.  

For the N-F-SC stacking, the F layer effectively acts as a potential barrier between the N and the SC layers. This strongly limits the strength of the proximity-induced superconductivity, unless the thickness of the F layer is comparable to the penetration depth of the modes into the ferromagnetic insulator. It appears unlikely that the F layer in the experiment is sufficiently thin that this mechanism would be operative. 

Our approach neglects various aspects of the experimental hybrid nanowires, such as the detailed device geometry, disorder, or effects of band bending. Nevertheless, we find that it is the SC-N-F arrangement which exhibits the most robust topological superconducting phase and provides the most likely Majorana-based explanation for the observation of zero-bias peaks in Ref.\ \cite{Vaitieknas2020}.\\

\begin{acknowledgments}
We gratefully acknowledge discussions with C.M.\  Marcus which initiated this work and funding by Deutsche Forschungsgemeinschaft through project C03 of CRC 183.
\end{acknowledgments}

%

%\bibliography{references}

\end{document}